\documentclass[11pt,fleqn,twoside]{article}
\usepackage{latexsym}
\makeatletter
\newcommand{\prava}{\footnotesize\it
\begin{flushright}
\begin{minipage}{6cm}
Copyright \copyright 1998 by M.L. Gandarias and  M.S. Bruzon
\end{minipage}
\end{flushright}}

\newcommand{\name}[1]{\begin{flushleft}
                       \LARGE \bf #1
                       \end{flushleft}\vspace{-3mm}}

\newcommand{\Author}[1]{\begin{flushleft}
                       \it #1 \end{flushleft}}

\newcommand{\Adress}[1]{\begin{flushleft}
                       \it #1 \end{flushleft}}

\newcommand{\Date}[1]{\begin{flushleft}
                      \small  \it #1 \end{flushleft}}

\newcommand{\ehkol}{Author \ name}
\newcommand{\ohkol}{Article \ name}
\renewcommand{\@evenhead}{
\hspace*{-3pt}\raisebox{-15pt}[\headheight][0pt]{\vbox{\hbox to \textwidth
{\thepage \hfil \ehkol}\vskip4pt \hrule}}}
\renewcommand{\@oddhead}{
\hspace*{-3pt}\raisebox{-15pt}[\headheight][0pt]{\vbox{\hbox to \textwidth
{\ohkol \hfil \thepage}\vskip4pt\hrule}}}
\renewcommand{\@evenfoot}{}
\renewcommand{\@oddfoot}{}

\newcommand{\be}{\begin{equation}}
\newcommand{\ee}{\end{equation}}
\newcommand{\ba}{\hspace*{-5pt}\begin{array}}
\newcommand{\ea}{\end{array}}

\newcommand{\ds}{\displaystyle}
\makeatother

\begin{document}
\setcounter{page}{8}

\thispagestyle{empty}

\renewcommand{\ehkol}{M.L. Gandarias and  M.S. Bruzon}
\renewcommand{\ohkol}{Symmetries of a
Generalized Boussinesq Equation}

\begin{flushleft}
\footnotesize {\sf
Journal of Nonlinear Mathematical Physics \qquad 1998, V.5, N~1},\ 
\pageref{gandarias_bruzon-fp}--\pageref{gandarias_bruzon-lp}.
\hfill
{{\sc Letter}}
\end{flushleft}

\vspace{-5mm}

{\renewcommand{\footnoterule}{}
{\renewcommand{\thefootnote}{}  \footnote{\prava}}

\name{Classical and Nonclassical Symmetries of \\
a Generalized Boussinesq Equation} \label{gandarias_bruzon-fp}

\Author{M.L. GANDARIAS and  M.S. BRUZON}

\Adress{Departamento de Matematicas, Universidad de Cadiz, \\
P.O. BOX 40, 11510, Puerto Real, Cadiz, Spain\\
E-mail: mlgand@merlin.uca.es}

\Date{Received August 19, 1997; Accepted October 30, 1997}

\begin{abstract}
\noindent
We apply the Lie-group formalism and the nonclassical method due to
Bluman and Cole to deduce symmetries of the generalized Boussinesq
equation, which has the classical Boussinesq equation as an special
case. We study the class of functions $f(u)$ for which this equation
admit either the classical or the nonclassical method. The reductions
obtained are derived. Some new exact solutions can be derived.
\end{abstract}


\strut\hfill

\noindent
The Boussinesq equation arises in several physical applications, the
f\/irst one was propagation of long waves in shallow water
\cite{gandarias_bruzon:bo}.
There have been several generalizations of the Boussinesq equation
such as the modif\/ied Boussinesq equation, or the dispersive water
wave.

Another  generalized Boussinesq equation is
\begin{equation}\label{gandarias_bruzon:ed1}
u_{tt}-u_{xx}+(f(u)+u_{xx})_{xx}=0,
\end{equation}
which has the classical Boussinesq equation as an special case for
$\ds f(u)=\frac{u^2}{2}+u$.
Recently conditions for the f\/inite-time blow-up of solutions of
(\ref{gandarias_bruzon:ed1}) have been investigated by Liu
\cite{gandarias_bruzon:liu}.

In this work we classify the Lie symmetries of
(\ref{gandarias_bruzon:ed1}) and we
study the class of functions $f(u)$ for which this equation is
invariant under a Lie group of point transformations.
Most of the required theory and description of the method can be
found in
\cite{gandarias_bruzon:bluku1,gandarias_bruzon:olv,gandarias_bruzon:ovsi}.

Motivated by the fact that symmetry reductions for many PDE's are
known that are not obtained using the classical Lie group method,
there have been several generalizations of the classical Lie group
method for symmetry reductions. Clarkson and Kruskal
\cite{gandarias_bruzon:cla}
introduced an algorithmic method for f\/inding symmetry reductions,
which is known as the direct method.
Bluman and Cole \cite{gandarias_bruzon:bluu} developed the nonclassical method to
study the symmetry reductions of the heat equation. The basic idea of
the method is to require that both the PDE (\ref{gandarias_bruzon:ed1}) and the
surface condition
\pagebreak

\begin{equation}\label{gandarias_bruzon:sur}
\Phi \equiv p\frac{\partial u}{\partial x}+q \frac{\partial
u}{\partial t}-r=0,
\end{equation}
must be invariant under the inf\/initesimal generator.
These methods were generalized and called conditional symmetries by
Fushchych {\em et al} \cite{gandarias_bruzon:fu} and also by Olver
and Rosenau
\cite{gandarias_bruzon:OR1, gandarias_bruzon:OR2}
to
include 'weak symmetries', 'side conditions' or 'dif\/ferential
constraints'.

}


We consider the classical Lie group symmetry analysis of equation
(\ref{gandarias_bruzon:ed1}). Invariance of equation
(\ref{gandarias_bruzon:ed1}) under a Lie group of
point transformations with inf\/initesimal generator
\begin{equation}\label{gandarias_bruzon:vect}
V=p(x,t,u) \frac{\partial}{\partial x}+q(x,t,u)
\frac{\partial}{\partial t}+r(x,t,u) \frac{\partial}{\partial u}
\end{equation}
leads to a set of twelve determining equations for
the inf\/initesimals. For totally arbitrary $f(u)$, the only simmetries
are the group of space and time translations which are def\/ined by the
inf\/initesimal generators
\begin{equation}\label{gandarias_bruzon:ge1}
V_1=\frac{\partial}{\partial x}, \qquad V_2=\frac{\partial}{\partial t}.
\end{equation}
 We obtain travelling wave reductions
\begin{equation}\label{gandarias_bruzon:ma1}
z=x-\lambda t, \qquad  u=h(z),
\end{equation}
where $h(z)$, after integrating twice with respect to $z$, satisf\/ies
\begin{equation}\label{gandarias_bruzon:ee1}
h''+(\lambda^2-1)h+f(h)=k_1z+k_2,
\end{equation}
with $k_1$ and $k_2$ arbitrary constants.
The only functional forms of $f(u)$, with $f(u)\neq \mbox{const.}$ and $f(u)$
nonlinear, which have extra symmetries are given in Table 1

\strut\hfill

\begin{tabular}{|lll|}
\multicolumn{3}{l}{
{\bf Table 1: }{ Symmetries for the generalized Boussinesq
equation.}} \\[3pt]
 \hline
\mbox{ }  & \mbox{ }  & \mbox{ }   \\[-10pt]
$i$ & $f(u)$ & $V^{i}_3$ \\
\mbox{ }  & \mbox{ }  & \mbox{ }   \\[-10pt] \hline
\mbox{ }  & \mbox{ }  & \mbox{ }   \\
$1$ & $d(au+b)^n+u+c$ & $\ds x \frac{\partial}{\partial x} +2t \frac{\partial}{\partial t}+ \frac{2}{a(1-n)}(au+b)\frac{\partial}{\partial u}$\\
\mbox{ }  & \mbox{ }  & \mbox{ }    \\
$2$ & $d\log(au+b)+u+c$ & $\ds x\frac{\partial}{\partial x}+2t\frac{\partial}{\partial t}+\frac{2}{a}(au+b)\frac{\partial}{\partial u}$\\
\mbox{ }  & \mbox{ }  & \mbox{ }   \\
$3$ & $d e^{(au+b)}+u+c$ & $\ds x\frac{\partial}{\partial x}+2t\frac{\partial}{\partial t}-\frac{2}{a}\frac{\partial}{\partial u}$\\
\mbox{ }  & \mbox{ }  & \mbox{ }   \\
\hline
\end{tabular}

\strut\hfill

We observe that equation (\ref{gandarias_bruzon:ee1}) with $f(h)=d(ah+b)^n+kh$ can
be solved.

Setting $\lambda^2=1-k$ and $k_1=0$  the solution is
\begin{itemize}
\item
For $m=n+1$ and $n \neq -1,$
\[
\pm\left (\frac{am}{2}\right
)^{1/2}\int{(-a(k_2h+k_3)m-d(ah+b)^m)^{-1/2}dh}=z+k_4.
\]
\item
For $n=-1,$
\[
\pm\left(\frac{a}{2}\right)^{1/2}\int{(-a(k_2h+k_3)-d\, \mbox{log
}(ah+b))^{-1/2}dh}=z+k_4.
\]
\item
For $n=2$, depending upon the choice of the constants, this equation is
solvable; for $k_1 \neq 0$ in terms of the f\/irst Painlev\'e equation,
and elliptic or elementary functions if $k_1=0.$
\item
For $n=3$, setting $k_1=k_2=0$, the equation is solvable in terms of
the Jacobi elliptic functions.
\end{itemize}

In Table 2 we list the corresponding similarity variables
and similarity solutions.

\strut\hfill

\begin{tabular}{|lllll|}
\multicolumn{5}{p{11cm}}{
{\bf Table 2:} {Each row gives the functions $f(u)$ for
which (\ref{gandarias_bruzon:ed1}) can be reduced to an ODE, as well as the
corresponding similarity variables and similarity
solutions.}}\\[20pt]
 \hline
\mbox{ } & \mbox{ } & \mbox{ }  & \mbox{\hspace{1.5cm} } & \mbox{ }\\[-10pt]
$i$ & $V^{i}_{3}$ & $f(u)$ & $z_i$ & $u_i$   \\
\mbox{ } & \mbox{ } & \mbox{ }  & \mbox{ } & \mbox{ }\\[-10pt]
\hline
\mbox{ } & \mbox{ } & \mbox{ }  & \mbox{ } & \mbox{ }\\
$1$ & $ V^{1}_3$ & $f=d(au+b)^n+ku+c$ &  $\ds \frac{x}{\sqrt{t}}$ &
$\ds t^{\frac{1}{1-n}}h(z)-\frac{b}{a} $    \\
\mbox{ } & \mbox{ } & \mbox{ }  & \mbox{ } & \mbox{ }\\
$2$ & $ V^{2}_3$ & $f=d\log(au+b)+u+c$ &  $\ds \frac{x}{\sqrt{t}}$ &
$\ds t h(z)-\frac{b}{a} $   \\
\mbox{ } & \mbox{ } & \mbox{ }  & \mbox{ } & \mbox{ }\\
$3$ & $ V^{3}_3$ & $f=de^{(au+b)}+u+c$ &  $\ds \frac{x}{\sqrt{t}}$ &
$\ds -\frac{1}{a}\log(t h(z)) $    \\
\mbox{ }  & \mbox{ }  & \mbox{ } & \mbox{ }& \mbox{ }  \\ \hline
\end{tabular}

\strut\hfill

In the Table 3 we show the ODE's to which PDE
(\ref{gandarias_bruzon:ed1}) is reduced by

\strut\hfill

\begin{tabular}{|ll|}
\multicolumn{2}{l}{
{\bf Table 3: }{ Symmetries for the generalized Boussinesq
equation with $k=nda^n.$ }}\\[3pt]
\hline
\mbox{ }  & \mbox{ }   \\[-10pt]
$V^{i}_3$ & $ODE i$\\
\mbox{ }  & \mbox{ }   \\[-10pt]
\hline
\mbox{ }  & \mbox{ }   \\
$V^1_3$ & $\ds h''''+\left(\frac{z^2}{4}+kh^{n-1}\right)h''+
k(n-1)h^{n-2}(h')^2+\left(\frac{z}{n-1}+\frac{3z}{4}\right)h'+\frac{nh}{(n-1)^2}=0$\\
\mbox{ }  & \mbox{ }      \\
$V^2_{3}$ & $\ds 4h^2h''''+4d(hh''-(h')^2)+h^2(z^2h''-zh')=0$\\
\mbox{ }   & \mbox{ }   \\
$V^3_3$ & $4gg'''+z^2(g')^2+2z^2-zg+k_1z+k_2-d e^{-g'}=0$\\
\mbox{ }   & \mbox{ }   \\
\hline
\end{tabular}

\strut\hfill

\begin{itemize}
\item
ODE1 for $n=2$, multiplied by $z$, can be integrated once and we
obtain
\[
\left({{z^{3}}\over{4}}+h\,k\,z\right)h'+h\,z^{2}
 +\,z h'''-{{h^{2}\,k}\over{2}}-h''=0.
\]
\item
ODE1 for $n=3,$ can be integrated once and we obtain
\[
 \left({{z^{2}}\over{4}}+h^{2}\,k\right)h'+{{3\,h
 \,z}\over{4}}+h'''=0.
\]
\item
ODE1 for $n=-1$, integrating once, we obtain
\[
 \left({{z^{2}}\over{4}}+{{k}\over{h^{2}}}
 \right)h'-{{h\,z}\over{4}}+h'''=0.
\]
\item
ODE3 with $h=e^{g'}$ has been obtained after integrating twice with
respect to $z.$
\end{itemize}


In the nonclassical  method one requires only the subset of solutions of
(\ref{gandarias_bruzon:ed1}) and (\ref{gandarias_bruzon:sur}) to be invariant under the inf\/initesimal
generator (\ref{gandarias_bruzon:vect}). In the case $q \neq 0$ we may
set $q(x,t,u)=1$ in (\ref{gandarias_bruzon:sur}
without loss of
generality. The
nonclassical method applied to (\ref{gandarias_bruzon:ed1}) gives rise to a set of
eight nonlinear determining equations for the inf\/initesimals.
The solutions for these equations depend on the function $f(u)$.
We can distinguish the following cases:
For
\[
f(u)=du^2+bu+c,
\]
 the solution is
\[
p=-d(p_1(t)x+p_2(t)),
\]
\[
r=p_1(p'_1+2p_1^2)x^2+(p_1p'_2+p_2p'_1+4p_1^2p_2)x+2dp_1u+
p_2p'_2+2p_1p_2^2+(1-b)p_1,
\]
where
$\ds p_1(t)=\frac{h'(t)}{2h}$, $\ds p_2(t)=
k_1p_1\int{\frac{h(t)dt}{(h'(t))^2}}+k_2p_1$,
and $h(t)$ satisf\/ies
\begin{equation}\label{gandarias_bruzon:eq}
(h'(t))^2=k_3h^3+k_4
\end{equation}
Here $k_1,\ldots,k_4$ are arbitrary costants.
Equation (\ref{gandarias_bruzon:eq}) is solvable in terms of the Weierstrass elliptic
functions if $k_3k_4 \neq 0$, and in terms of elementary functions otherwise.
Solving (\ref{gandarias_bruzon:eq}) for $h(t)$ yields the six
canonical symmetry
reductions derived by Clarkson for the classical
Boussinesq equation using the nonclassical method, and by Clarkson
and Kruskal using the direct method.

For any other function $f(u)$ listed in Table 1 the same
symmetries, as were obtained by the classical method, appear.

 \label{gandarias_bruzon-lp}
\end{document}